%%%%%%%%%%%%%%%%%%%%%%%%%%%%%%%%%%%%%%%%%%%%%%%%%%%%%%%%%%%%%%%%%%%%%%%%
%%                TOP OF FILE - DESCRIPTOR                            %%
%%%%%%%%%%%%%%%%%%%%%%%%%%%%%%%%%%%%%%%%%%%%%%%%%%%%%%%%%%%%%%%%%%%%%%%%
%
%% @texfile{
%%     filename=as_article.tex",
%%     date=OCT-23-1997",
%%     filetype="Euclidean Reconstruction in Quantum Field Theory:
%%               Between tempered distributions and Fourier Hyperfunctions",
%%     author=Andreas U. Schmidt",
%%     address="Johann Wolfgang Goethe-Universit{\"a}t,
%%              D--60054 Frankfurt am Main",
%%     email="Internet: aschmidt@math.uni-frankfurt.de",
%%     codetable="ISO/ASCII",
%%     keywords="quantum field theory, schwinger functions,
%%               wightman functions, euclidean field theory, hyperfunctions",
%%     abstract="This file is an extended writeup of an talk held on the
%%               second Operator Theory Seminar Frankfurt-Krakow held in
%%               Krakow, April 1997"
%%     }
%
%%%%%%%%%%%%%%%%%%%%%%%%%%%%%%%%%%%%%%%%%%%%%%%%%%%%%%%%%%%%%%%%%%%%%%%%
%%                          MACRO DEFINITIONS                         %%
%%%%%%%%%%%%%%%%%%%%%%%%%%%%%%%%%%%%%%%%%%%%%%%%%%%%%%%%%%%%%%%%%%%%%%%%
%
% --- generic style defnitions for the journal ---
%
\input amstex
\documentstyle{amsppt}
\Monograph
\loadbold

\NoRunningHeads
%\NoPageNumbers
\catcode`\@=11
\def\logo@{}
\def\proclaimheadfont@{\smc}
\def\demoheadfont@{\smc}

\def\output@{\shipout\vbox{%
 \iffirstpage@ \global\firstpage@false
   \hbox to\hsize
    {\hfil\sixrm
     %UNIVERSITATIS IAGELLONICAE ACTA MATHEMATICA, FASCICULUS XXXIV
     \hfil}
   \vskip 2pt
   \hrule
  \vskip 4pt
  \hbox to\hsize{\sixrm\hfil April 1997 \hfil}
  %\vskip 0.5truecm
  \pagebody %\logo@
%  \makefootline%
 \else
     \hbox to\hsize{{\tenrm\ifodd\pageno \hss\folio \else \folio\hss \fi}}
   \vskip 0.7truecm
%
%  \ifrunheads@ \makeheadline
   \pagebody
% \makefootline
 \fi}%
 \advancepageno \ifnum\outputpenalty>-\@MM\else\dosupereject\fi}

\catcode`\@=\active

\magnification=\magstephalf

\baselineskip 18pt

\hsize 13 truecm
\vsize 19 truecm

\LimitsOnSums
\LimitsOnNames
\CenteredTagsOnSplits

\def\addr#1{{\eightpoint\phantom{.}\hskip 7.5truecm#1 \newline}}

\def\theo#1{\procla{Theorem #1} }
\let\endtheo\endproclaim

\def\lem#1{\procla{Lemma #1} }
\let\endlem\endproclaim

\def\proof#1{\demol{Proof #1}}
\let\endproof\enddemo

\def\prop#1{\procla{Proposition #1}}
\let\endprop\endproclaim

\let\demol\demo
\let\procla\proclaim

\define \bo {\bold}
\define \oti {\otimes}

\define\pc#1#2{ \frac {\partial {#1}}{\partial {#2}} }
\define\pcc#1#2{ \frac {\eth {#1}}{\eth {#2}} }
\define\pct#1#2#3{ \frac {\partial^{#1} {#2}}{\partial {#3}^{#1}} }
\define\pzt#1#2#3{ \frac {d^{#1} {#2}}{d{#3}^{#1}} }
\define\pz#1#2{ \frac {d{#1}}{d{#2}} }

%
% --- the author's definitions start here ---
%
\loadeusm % --- I extesively use that font for math symbols
%
%   --- definitions of symbols ---
%

\def\MM{\mathord{\eusm M}}

\def\forw{\mathord{\overline{V}^{\,+}}}
\def\SM{\mathord{\eusm S(\eusm M)}}

\def\SB{\mathord{\underline{\eusm S}}}

\def\SR#1{\mathord{\eusm S(\bold R^{#1})}}

\def\supp{\mathord{\roman{supp}}}
\def\WW{\mathord{\frak W}}
\def\Wf{\mathord{\bold W}}
\def\SS{\mathord{\frak S}}
\def\Sf{\mathord{\bold S}}
%
 % --- fraktur symbols for spaces

%
%\def\SR{\mathord{{\eusm S}(\bold R)}}

\def\Im{\mathop{\text{Im}}}
\def\Re{\mathop{\text{Re}}}
\def\OM{\mathop{\eusm O_c^m}}
\def\OO{\mathord{\eusm O}}
\def\PII{\mathord{{\eusm P}_{\ast\ast}}}
\def\PIIp{\mathord{{\eusm P}^\prime_{\ast\ast}}}
\def\PI{\mathord{{\eusm P}_\ast}}
\def\PIp{\mathord{{\eusm P}^\prime_\ast}}
\def\PR{\mathord{{\eusm P}_\rho}}
\def\PRp{\mathord{{\eusm P}^\prime_\rho}}
\def\OII{\mathop{\eusm O}\limits^\approx}
\def\SAB{\mathord{\eusm S_\alpha^\beta}}
\def\SA{\mathord{\eusm S_\alpha}}
\def\SB{\mathord{\eusm S^\beta}}
%
%   --- abbreviative macros ---
%
\define\thismonth{\ifcase\month % case 0 --- impossible!
  \or January\or February\or March\or April\or May\or June%
  \or July\or August\or September\or October\or November%
  \or December\fi}
%

%

%
 % equation numbers and references for
          % the appendix
%

%

%
\def\diag{\text{diag}}
%
%%%%%%%%%%%%%%%%%%%%%%%%%%%%%%%%%%%%%%%%%%%%%%%%%%%%%%%%%%%%%%%%%%%%%%%%
%%                           TITLE PAGE                               %%
%%%%%%%%%%%%%%%%%%%%%%%%%%%%%%%%%%%%%%%%%%%%%%%%%%%%%%%%%%%%%%%%%%%%%%%%
%
\topmatter
\title
Euclidean Reconstruction in Quantum Field Theory:\\
Between tempered distributions and Fourier Hyperfunctions
\endtitle
\author
by Andreas U. Schmidt\\
\endauthor
\abstract\nofrills{{\bf Abstract.}}
In this short note on my talk I want to point out the mathematical
difficulties that arise in the study of the relation of Wightman and
Euclidean quantum field theory, i.e., the relation between the hierarchies
of Wightman and Schwinger functions. The two extreme cases where the
reconstructed Wightman functions are either tempered distributions --- the
well-known Osterwalder-Schrader reconstruction --- or modified Fourier
hyperfunctions are discussed in some detail. Finally, some perpectives
towards a classification of Euclidean reconstruction theorems are outlined
and preliminary steps in that direction are presented.
\endabstract
\endtopmatter
\document
%
%%%%%%%%%%%%%%%%%%%%%%%%%%%%%%%%%%%%%%%%%%%%%%%%%%%%%%%%%%%%%%%%%%%%%%%%%
%%                               MAIN TEXT                             %%
%%%%%%%%%%%%%%%%%%%%%%%%%%%%%%%%%%%%%%%%%%%%%%%%%%%%%%%%%%%%%%%%%%%%%%%%%
%
% -------------------------------------------
\head 0.~Introduction: Why Euclidean Reconstruction \endhead
Euclidean methods are widely used in quantum field theory and other areas
of mathematical physics. So let me outline some of the reasons for the
attractivity of those techniques. I will use the language of axiomatic
``Wightman'' quantum field theory, for the fundamental notions of which I
may refer the reader to my last years talk~\cite{1} or the famous
books~\cite{SW} of Streater and Wightman or~\cite{GJ} of Glimm and Jaffe
respectively.

To use ``Euclidean'' methods in quantum field theory amounts formally in
changing the coordinates $x=(x^0,x^1,x^2,x^3)=(t,\vec{x})$ of Minkowski space
to $\xi=\imath {}^tx=(it,\vec{x})$, where $\imath=(i,1,1,1)$. By that the original
Minkowski metric $\eta=\diag(-1,1,1,1)$ prescribed by relativistic covariance
of the theory becomes the trivial metric of Euclidean space, but the new time
coordinate is purely imaginary. The benefits of this transformation is
somewhat na\"{\i}vely but strikingly explained in terms of the {\it path
integral}. This is a formal expression for the so called
{\it generating functional} of a quantum field theory which reads
$$
\eusm Z\{f\}=\int\eusm D\Phi\, e^{\displaystyle i \Phi(f)+i \eusm L(f)}.
$$
From this expression all {\it correlation functions} of the theory, i.e.,
actual probabilities of physical events, can in principle be derived by a
formal {\it variational derivative} of $\eusm Z\{f\}$ with respect to the test
function $f\in\SM$, which is usually assumed to be tempered, so that the
{\it field} $\Phi\in\SM'$ will be a tempered distribution. The exponent in
this expression, termed {\it action integral}, is divided into the {\it source}
part $\Phi(f)$ and the  {\it Lagrangean} $\eusm L(f)$ which is a
function (usually a polynomial) of the field $\Phi$ and its derivatives.
Following the principles
of variational theory we know that the {\it classical} field configurations,
i.e., solutions of the Euler-Lagrange equations derived from $\eusm L$,
minimize the action, so that their contribution to $\eusm Z\{f\}$ is maximal,
whereas fluctuations around the classical situation are supposed to give
smaller contributions. This is what essentially allows one to approximate
$\eusm Z\{f\}$ by perturbation expansions, e.g., in Planck's constant $\hbar$
(semiclassical approximation) or, most commonly, in a {\it coupling constant}
$g$ appearing in $\eusm L$, which leads to the vast field of {\it diagrammatic
perturbation theory}, see, e.g.,~\cite{tH}. The free, classical fields around
which this expansion takes place are
solutions of simple partial differential equations like the wave, Klein-Gordon,
and free Dirac equation, in that order:
$$
\square\phi=0,\quad
(\square+m^2)\phi=0,\quad
(i\gamma^\mu\partial_\mu-m)\phi=0
$$
(the classical field $\phi$ should not be mixed up with the operator valued
distribution $\Phi$). These equations follow as  Euler-Lagrange equations
from the {\it free} part $\eusm L_{\text{f}}$ of the Lagrangean
$\eusm L=\eusm L_{\text{f}}+\eusm L_{\text{i}}$.

The mathematical properties of $\eusm Z\{f\}$ are not very satisfactory: The
{\it path space measure} $\eusm D\Phi$ over the space of all possible
field configurations cannot be defined because of the unknown geometrical
structure of this space, and even if it would be well-defined, the integral
would by no means be guaranteed to converge. The situation changes dramatically
after the seemingly innocent change to Euclidean coordinates. The new
generating functional
$$
\frak S\{f\}=\int d\mu\, e^{\displaystyle -\Phi(f)-\eusm L(f)}
$$
is an integral with respect to a well-defined {\it Borel probability measure}
$d\mu$ on $\eusm S(\bold R^4)'$ and converges due to the negative
definite damping factor in the exponent (usually like a Gaussian).

On the level of correlation functions, which are derived from the generating
functionals or are the input of the axiomatic approach, the advantage
of working in the  Euclidean regime is reflected by the fact that the hierarchy
of Wightman $n$-point functions $\WW_n$, $\Wf_n$, are tempered distributions,
whereas their Euclidean equivalents --- the so-called {\it Schwinger functions}
--- are indeed {\it real analytic functions}.

So from a general point of view, we have turned a quantum theory into a theory
which is nothing but a formulation of {\it statistical mechanics} by a simple
change of coordinates. It is not surprising that very  commonly physicists use
the Euclidean framework to do actual calculations, transferring their results
back to the quantum regime by writing $t$ for $it$ again. However, somewhere
in this ``backtransformation'' the nontrivial quantum nature should reappear,
which means in terms of correlation functions that the Schwinger functions
should gain singularities in the process of analytic continuation back
to Minkowski spacetime. So, besides the mathematical justification of the use
of Euclidean methods, a consistent theory of what is called {\it Euclidean
reconstruction} of the Wightman hierarchy from the Schwinger hierarchy would
help to answer fundamental questions about quantum field theory in the
axiomatic framework. Let us first walk the simpler path, namely the continuation
of Wightman functions to the Euclidean domain.
%
% -------------------------------------------
\head 1.~Analytic continuation of Wightman
Functions to the Euclidean Domain\endhead
Starting from the Wightman functions (in difference variables)
$\Wf_{n+1}\in\SR{4n}$ which are distributions in
the real variables $x_1,\ldots,x_n$, $x_i=(x_i^0,\vec{x}_i)\in\bo R^4$, we want to
see that they can be extended to holomorphic functions on some domain in
$\bo C^{4n}$ and determine the shape of that domain. In this process, the physical
axioms imposed on the Wightman functions (see~\cite{1}) enter at different stages.

In the first step we use the fact that the Fourier transforms
$\widetilde{\Wf}_{n+1}$ have their support
in the $n$-fold product of the forward lightcone by the {\it spectrum condition}:
$\supp\widetilde{\Wf}_{n+1}\subset\forw^n$,
$\forw\equiv\{x\in\MM\mid (x,x)\geq0, x^0\geq0\}$, where $(x,x)={}^tx\eta x$
denotes the usual
Minkowski inner product. By that fact and a general theorem on the Laplace
transformation of tempered distributions, see, e.g.,~\cite{RS}, theorem~IX.16,
$\Wf_{n+1}$ are boundary values in
the sense of $\SR{4n}'$ of holomorphic functions in the {\it forward tube}
$T_n\equiv\bo R^{4n}+i{V^{+}}^n$. This theorem is a generalization of the famous
Payley-Wiener-Schwartz theorem for distributions with compact support to the
tempered case. The characterizing property of the Fourier-Laplace transforms
of tempered distributions with support in some cone is that they are
polynomially bounded. This will be of crucial importance in the Euclidean
reconstruction of tempered distributions, as we will see below.

Next one uses the {\it invariance}
 of the Wightman functions under the action of
the {\it Lorentz group}. This is the connected component $\text{L}(\bold R)$
containing the identity
of
$$
\{g\in\text{GL}(4,\bold R)\mid (gx,gx)=(x,x)\text{ for }x\in\bold R^4\},
$$
where the action of $\text{L}(\bold R)$
 is extended to $\bold R^{4n}$ in the obvious way.
The key is the following
\theo{[Bargmann-Hall-Wightman]}
If a function $\Wf_{n+1}$ is holomorphic on the tubular domain $T_n$
and invariant
under the action of the Lorentz group then $\Wf_{n+1}$ can be analytically
continued to a single valued function on the {\rm extended tube}
$$
T_n'\equiv\bigcup_{g\in\text{L}(\bold C)}gT_n\subset\bold C^{4n},
$$
where $\text{\rm L}(\bold C)$ denotes the {\rm complexification} of the
Lorentz group, and $\Wf_{n+1}$ is invariant under the action of
$\text{\rm L}(\bold C)$.
\endtheo
This may be found, e.g., in~\cite{SW}, theorem~2.11. The difficult point in
its proof is the single-valuedness of the analytic continuation.

Unlike the tubular domain $T_n$, the extended tube $T_n'$ contains real
points. These points have the remarkable property of being alltogether
{\it spacelike}: We call $x\in\bold R^4$ {\it spacelike} if $(x,x)<0$,
{\it timelike} if $(x,x)>0$ and finally
$x=(x_1,\ldots,x_n)\in\bold R^{4n}$ is called a
 {\it Jost point} if it has the
property that $\lambda_1 x_1+\cdots+\lambda_n x_n$ is spacelike for any
$\lambda_j\geq 0$ with $\sum_{j=1}^n\lambda_j>0$. A fundamental lemma
due to Jost is
\lem{[Jost]}
The set $T_n'\cap\bold R^{4n}$ is open and we have
$$
T_n'\cap\bold R^{4n}= \text{ the totality of Jost points of }\bold R^{4n}.
$$
\endlem
An accessible proof may be found in~\cite{MO}, chapter~9,~\S6.
Combining this with the above theorem we find that the $\Wf_{n+1}$ are real
analytic functions on the set of Jost points.

Now remember that the $\Wf_{n+1}$ in the difference variables are defined by
the Wightman functions $\WW_{n+1}$ in the ordinary variables,
for which we simply use the letter $y$, by
$$
\WW_{n+1}(y_1,\ldots,y_{n+1})=\Wf_{n+1}(y_2-y_1,\ldots,y_{n+1}-y_{n}),
$$
and that {\it causality} is imposed on the $\WW_{n+1}$:
$$
\WW(y_1,\ldots,y_j,y_{j+1},\ldots,y_{n+1})=
\WW(y_1,\ldots,y_{j+1},y_j,\ldots,y_{n+1}),
$$
whenever $x_j=y_j-y_{j+1}$ is spacelike. Now in the Jost points all differences
are spacelike so that we can apply any permutation to the $n+1$ variables
$y_1,\ldots,y_{n+1}$ and will obtain the same analytic function $\Wf_{n+1}$
when going back to difference variables. So, finally the $\Wf_{n+1}$ can be
uniquely extended to the {\it permuted extended tube} defined by
$$
{T_n'}^\Pi\equiv \{ \pi x \mid x \in T_n',\ \pi\in S_{n+1} \},
$$
where for $x=(y_2-y_1,\ldots,y_n-y_{n-1})$ and a permutation
$\pi\in S_{n+1}$  we define
$\pi x\equiv(y_{\pi(2)}-y_{\pi(1)},\ldots,y_{\pi(n+1)}-y_{\pi(n)})$.

The last step will be to show that ${T_n'}^\Pi$ contains the Euclidean points.
Let us first state what we mean by these: A point
$x=(y_2-y_1,\ldots,y_{n+1}-y_{n})\in\bold C^{4n}$ is called
{\it Euclidean point} if it is of the form
$(\imath y_2'-\imath y_1',\ldots,\imath y_{n+1}'-\imath y_{n}')$, with
$(y_1',\ldots,y_{n+1}')\in\bold R^{4(n+1)}$, i.e., it has purely imaginary time
and purely real space components. An Euclidean point is called
{\it non-coincident} if $y_j\neq y_k$ for all $j\neq k$.
\prop{}
The permuted extended tube ${T_n'}^\Pi$ contains all the non-coin\-cident
Euclidean points of $\bold C^{4n}$.
\endprop
We sketch the proof following~\cite{BO}, proposition~9.10:
{\proof{}
Begin with the simplest situation, where all the time components $y_j^{\prime
0}$ are
distinct for $1\leq j \leq n+1$. Then, by a suitable permutation $\pi$ which
arranges the $y_j'$ such that $y_{\pi(j)}^{\prime 0}$ increases when $j$
increases,
we find that the difference vector $x$ determined by $\imath \pi y'$ lies in
$T_n$ since
$y_{\pi(j)}^{\prime 0}-y_{\pi(j+1)}^{\prime 0}>0$,
which means that $\imath$ applied to all these difference vectors lies
in the forward tube.
In the general situation we may nevertheless assume without loss of generality
$y_1^{\prime 0}\leq\ldots\leq y_{n+1}^{\prime 0}$, by applying a suitable
permutation if necessary. Consider the collection of vectors in $\bold R^3$
$$
\{e_{jk}\equiv y_j'-y_k'
\mid y_j^{\prime0}=y_k^{\prime0},\ j,k=1,\ldots,n+1,\ j\neq k\}.
$$
Since all these vectors are non-zero on non-coincident points we may find a
three-dimensional unit vector $\vec{s}$ that is not orthogonal to any of them.
Assume, again without loss of generality, that $\vec{s}$ is a vector along the
$y^{\prime 3}$-axis. Then we may well-order the set $\{y_j'\}$ according
to their $y^{\prime 3}$-component. Now the Lorentz-transformation
$$
\Lambda_\beta=\left(\matrix \cos\beta   & 0 & 0 & \sin\beta \cr
                                0       & 0 & 0 &      0     \cr
                                0       & 0 & 0 &      0     \cr
                            \sin\beta & 0 & 0 &  \cos\beta \cr \endmatrix\right)
$$
rotates the difference vector $x$ corresponding to $\imath y'$
into $T_n\subset {T_n'}^\Pi$ for suitable $\beta$.
\endproof}
We call the restriction of the Wightman functions to the non-coincident
Euclidean region the {\it Schwinger functions}
of Euclidean quantum field theory and denote them by
$\SS_n$ in the ordinary and $\Sf_n$ in the difference variables.
As we have shown above, the $\eusm \SS_n$ are symmetric in their arguments, so
that it is enough to consider the $\Sf_n$ in the difference variables
as functions on
$\imath\bo R_+^{4(n-1)}\equiv (i\bo R_+\times\bo R^3)^{n-1}$:
$$
\Sf_n\equiv\Wf_n|_{\imath\bo R_+^{4(n-1)}}, \quad
\SS_n\equiv\WW_n(\imath y_1,\ldots,\imath y_n).
$$

Let us finally list the properties of the Schwinger functions:
{\parindent=1.5cm
 \item{$\bullet$} Real analyticity,
 \item{$\bullet$} Euclidean, i.e., $\text{SO}_4$-covariance,
 \item{$\bullet$} positivity, an equivalent of axiom B in~\cite{1}, and
 \item{$\bullet$} total symmetry.\par}
Further axioms on the relativistic side, like the {\it cluster
decomposition property} (axiom~C in~\cite{1}), will in general also
be reflected by counterparts on the Euclidean side, but see~\cite{SI}
for a deeper discussion.
%
% -------------------------------------------
\head 2.~Reconstruction of Tempered Distributions\endhead
The reconstruction of Wightman functions as tempered distributions was done by
Osterwalder and Schrader in~\cite{2}. Although the basic idea --- to use the Euclidean
covariance of the Schwinger functions to continue them analytically from their original
domain of definition $\bo C_+^{4n}$ --- is quite simple, the actual reconstruction
of tempered distributions turned out to be a great effort. The point is the following:

In the final step of reconstruction a Payley-Wiener-Schwartz-like theorem has to be
used to yield the Fourier transforms of the Wightman functions as boundary
values in the sense of $\eusm S'$ of the Fourier-Laplace transforms of the
analytically continued Schwinger functions. As we have already stated above, for such a
theorem to be applicable, the Schwinger functions must satisfiy certain polynomial
growth conditions. The difficulty lies in controling these growth conditions
in the process of analytic continuation. Let us sketch this somewhat formally:

The Schwinger function $\Sf_n$ is a  real analytic function on
$\imath\bo R_+^{4(n-1)}$ and
satisfies a {\it real estimate} of the form
$$
|\Sf_n(\xi)|\leq A_n^{(0)}\cdot E^{(0)}\left(\xi_j^0,|\vec{\xi_j}|\right),
\quad \imath\xi\in\imath\bo R_+^{4(n-1)}.
\eqno{(\text{RE})}
$$
Indeed such an estimate is the basis for the {\it equivalence} of relativistic and
Euclidean theory: The continuations of the Wightman functions naturally satisfy
the polynomial bound following from the Payley-Wiener-Schwartz theorem. That is,
for the reconstruction of {\it tempered distributions}, $E^{(0)}$ has the special form
$$
E^{(0)}=\left[
\left(1+\max_{1\leq j\leq n-1} |\vec{\xi}_j|\right)
\left(1+\sum_{j=1}^{n-1}\xi_j^0\right)
\left(1+\sum_{j=1}^{n-1}(\xi_j^0)^{-1}\right)
\right]^{nt},
$$
for some $t>0$. The analytic continuation of the time variables back to the real axis
can unfortunately not proceed in all $n-1$ variables simultaneously. In fact, this was
erroneously assumed in the first version of Osterwalder-Schrader's proof~\cite{3},
which led to the refinement in~\cite{2}. The way they circumvented this difficulty was
purely geometrical: In an inductive process they continued $\Sf_n$ to {\it cones}
$\Gamma^{(N)}$ around the imaginary time axis after $N$ steps. They showed that this
cones have increasing opening angle
$\pi/2(1-2^{-N/2})$ in every coordinate. They then showed that these cones exhaust the whole ``upper half
plane''
$$
\bigcup_{N\in\bo N} \Gamma^{(N)}=\bo C_+^{4(n-1)}=
\left( \{\zeta\in\bo C\mid\Im \zeta>0\}\times\bo R^3\right)^{(n-1)},
$$
as shown in the following sketch:
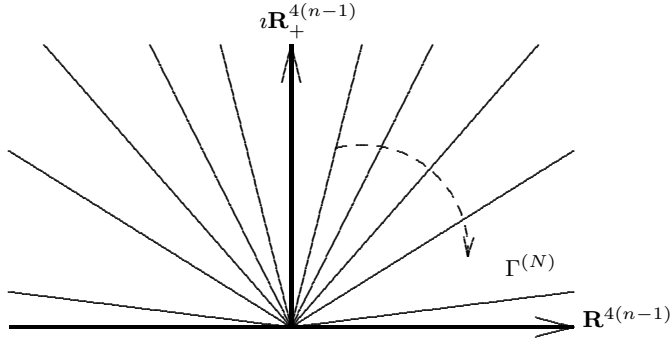
\input prepictex
\input pictex
\input postpictex
$$
\beginpicture
\eightpoint
\setcoordinatesystem units <.9cm,.9cm>
\linethickness=1pt
\setlinear
\putrule from  0.952 22.860 to  8.572 22.860
\plot  8.064 22.733  8.572 22.860  8.064 22.987 /
\putrule from  4.762 22.860 to  4.762 26.670
\plot  4.890 26.162  4.762 26.670  4.635 26.162 /
%\linethickness= 0.500pt
\plot  3.810 26.670  4.762 22.860 /
\plot  4.762 22.860  5.715 26.670 /
\setsolid
\plot  7.094 24.070  7.144 23.812  7.221 24.063 /
\setdashes < 0.1270cm>
\circulararc 112.620 degrees from  7.144 23.812 center at  5.715 23.892
\setsolid
\plot  6.668 26.670  4.762 22.860 /
\plot  4.762 22.860  2.858 26.670 /
\plot  0.952 23.336  4.762 22.860 /
\plot  4.762 22.860  8.572 23.336 /
\plot  1.429 26.670  4.762 22.860 /
\plot  4.762 22.860  8.096 26.670 /
\plot  8.572 25.241  4.762 22.860 /
\plot  4.762 22.860  0.952 25.241 /
\put {$\Gamma^{(N)}$} [cc] at 8.0 23.7
\put {$\bo R^{4(n-1)}$} [cc] at 9.3 23
\put {$\imath\bo R_+^{4(n-1)}$} [cc] at 5 27
\endpicture
$$
The continued Schwinger functions will then satisfy an estimate very much like (RE)
on $\Gamma^{(N)}$:
$$
  |S_n(\zeta)|\leq A_n^{(N)}\cdot
    E^{(N)}\left(|\Re\zeta_j^0|,|\Im\zeta_j^0|,|\vec{\zeta_j}|\right), \quad
  \zeta\in \Gamma^{(N)},
\eqno{(\text{NE})}
$$
where $E^{(N)}$ does not differ essentially from $E^{(0)}$, i.e., is also of
polynomial form. The difficulty is
now that the growth of the constant $A_n^{(N)}$ has to be controled in order
that an estimate like (NE) can hold on whole $\bo C_+^{4(n-1)}$, to make
an inverse Payley-Wiener-Schwartz theorem applicable. The solution of
Osterwalder-Schrader is to impose an additional condition on the original Schwinger
functions, the so-called {\it linear growth condition}:
$$
A_n^{(0)}\leq \alpha(n!)^\beta,
\eqno{(\text{LG})}
$$
for some constants $\alpha$, $\beta>0$\footnote{Notice
that in the axiomatic approach to
reconstruction the Schwinger functions are not {\it a priori} assumed to be real
analytic functions but also only tempered distributions. So (LG) is formulated
in~\cite{2} as a modification of the usual temperedness axiom, not directly as
a condition for the real analyticity constant $A_n^{(0)}$.}.
From the linear growth condition, an estimate of $A_n^{(N)}$ follows in the $N$-th step:
$$
A_n^{(N)}\leq n^{\beta n}\cdot 2^{\beta nN},
$$
where the factor $n^{\beta n}$ stems directly from (LG). It is a great technical
achievement of Osterwalder-Schrader to eliminate $N$ from this estimate by purely
geometrical methods! The final form of the estimate is then
$$
A_n^{(N)}\leq C_n \leq a b^{n^2},
$$
for some $a$, $b>0$. This estimate exhibits extremly rapid growth in $n$, but this
is immaterial for the final step of reconstruction, because one is now in the
position to apply an theorem of Vladimirov, see~\cite{V}, pp.~235, which renders
the boundary values of the continued Schwinger functions on $\bo R^{4(n-1)}$ as
tempered distributions.

The drawbacks of introducing (LG) are obvious:
{\parindent=1.5cm
 \item{$\bullet$} It is merely a sufficient condition for reconstruction of Wightman
functions, so that one cannot claim to have full equivalence between the Euclidean
and relativistic frameworks. This is also reflected by the fact that
 \item{$\bullet$} the reconstructed Wightman functions fulfill additional growth conditions
of the form
$$
|\WW_n(f)|\leq\gamma\delta^{n^2}\cdot||f||,
$$
for some $\gamma$, $\delta>0$ and a certain norm on $\eusm S(\bo R^{4(n-1)})$.
 \item{$\bullet$} The linear growth condition was introduced {\it ad hoc}.\par}
It was conjectured already by Osterwalder-Schrader that one could overcome these
problems by leaving the tempered case and considering Euclidean reconstruction for
larger distribution classes. This is what will concern us now.
%
% -------------------------------------------
\head 3.~Reconstruction of Fourier Hyperfunctions\endhead
The mathematical problems of quantum field theory manifest themselves partially
in the singularity structure of the correlation functions, see, e.g.,~\cite{1}
and~\cite{ST}, if one tries to include interaction into the formalism.
It was early pointed out by Arthur~S.~Wightman, see~\cite{4} and~\cite{5}, that this
calls for a generalization of distribution theory and several suggestions in
this direction have been made, see again~\cite{5} and the concise review
in~\cite{6} and references therein.

One of the most successful choices was to use {\it hyperfunctions}, which were
invented by Mikio Sato in the late 50's and whose fundaments were
published in~\cite{7} and~\cite{8}, in a great effort to give
rigorous meaning to the notion of ``boundary values of holomorphic functions'',
cf. the introductory texts~\cite{KA} and~\cite{MO}. Later, also going back to
ideas of Sato, the theory was extended by Kawai in his master's thesis and~\cite{9}
to include Fourier transformation, termed {\it Fourier hyperfunctions}, and further
to (Fourier) hyperfunctions with values in a Hilbert resp. Fr\'echet space by Yoshifumi
Ito and Shigeaki Nagamachi, see~\cite{10},~\cite{11}, and~\cite{12} and references
therein.

The formulation of axiomatic quantum field theory in terms of Fourier hyperfunctions
was carried out by Nagamachi and Nogumichi Mugibayashi in their fundamental
papers~\cite{13},~\cite{14}, and~\cite{15} and by Erwin Br\"uning and Nagamachi
in~\cite{5}, see~\cite{16},~\cite{17}, and~\cite{18} for concrete physical
applications. We will here be mainly outline the results of~\cite{15}, where an
Euclidean reconstruction theorem for the so called {\it Fourier hyperfunctions
of type II}, which are now commonly termed {\it modified Fourier hyperfunctions} is
proven.
We can give here only a very coarse sketch of their theory, so the reader may be
referred to~\cite{19} and~\cite{20} for a thorough treatment.

There are generally two ways to define any kind of hyperfunction: The
{\it duality method} in the sense of Schwartz' distribution theory, emphasizing the
r\^{o}le of the test function spaces, and the {\it algebro-analytic method}, which
views hyperfunctions as boundary values of holomorphic functions. But before going
into details, let us fix the topology of the spaces on which our hyperfunctions will
live:

{\bf Radial Compactification of $\bo C^{\bo n}$.}
To ensure that the Fourier tranfsormation will act as an isomorphism
on the test function
spaces to be defined below, and by that on the Fourier hyperfunctions, we have to
impose what may be loosely speaking called ``conditions at infinity.'' This is only
consistently possible if we {\it compactify} the coordinate space in a certain
manner: Let $\bo D^n\equiv \bo R^n \sqcup S_\infty^{n-1}$ be the disjoint union of $\bo R^n$ with
the points ``at infinity'' in every direction. The {\it complexification}
$\bo Q^n$ is identified with $\bo D^{2n}$ and is equipped with the following topology:
For $x\in\bo C^n$ take the usual fundamental system  of open neighbourhoods.
For $x\infty\in S_\infty^{2n-1}$ take open neighbourhoods
$x\infty\in C_\infty\subset S_\infty^{2n-1}$ and form their disjoint union with all
translations of the cone  with vertex at the origin $C\subset\bo C^n$ that has opening
in the direction of $C_\infty$, i.e.,
$C\cap S^{2n-1}=C_\infty$ under the obvious identification of $S_\infty^{2n-1}$
with $S^{2n-1}$. As a special form for these neighbourhoods we may choose
$$
C_\delta\sqcup C_\delta\infty, \quad C_\delta\equiv
\{ z\mid |\Im z|<\delta(|\Re z|+1)\}.
$$
where $C_\delta\infty$ denotes the points ``at infinity'' of $C_\delta$.
We will immediately use these open sets to define test function spaces.

{\bf Modified Fourier hyperfunctions as distributions.}
Choose special neighbourhoods of the real axis $\bo D^n$ in $\bo Q^n$ defined by
$$
V_m \equiv U_m^n\sqcup {(U_m^n)}_\infty,\quad
U_m \equiv\{z\in\bo C|\,|\Im z|<(1+|\Re z|)/m\},
\eqno{(V_m)}
$$
as shown below:
$$
\beginpicture
\eightpoint
\setcoordinatesystem units <.9cm,.9cm>
\setlinear\setsolid
\circulararc 87.206 degrees from  1.429 26.194 center at  3.929 23.812
\plot  5.308 23.492  5.048 23.527  5.261 23.374 /
\setdashes < 0.1270cm>
\circulararc 136.809 degrees from  5.048 22.384 center at  4.822 22.955
\setsolid
\plot  3.022 23.457  3.048 23.717  2.903 23.499 /
\setdashes < 0.1270cm>
\circulararc 39.308 degrees from  3.048 23.717 center at  6.382 22.527
\circulararc 67.380 degrees from  7.430 24.289 center at  6.834 24.479
\setsolid
\plot  6.637 24.419  6.382 24.479  6.578 24.306 /
\setdashes < 0.1270cm>
\circulararc 140.115 degrees from  6.001 21.526 center at  5.656 23.072
\setsolid
\plot  0.559 23.564  0.476 23.812  0.432 23.554 /
\setdashes < 0.1270cm>
\circulararc 83.471 degrees from  0.476 23.812 center at  3.150 24.016
\setsolid
\plot  1.842 23.940  1.334 23.812  1.842 23.685 /
\linethickness= 1pt
\putrule from  1.334 23.812 to  8.192 23.812
\plot  7.684 23.685  8.192 23.812  7.684 23.940 /
\linethickness= 0.500pt
\plot  1.429 25.241  4.762 24.289 /
\plot  4.762 24.289  8.096 25.241 /
\plot  1.429 22.384  4.762 23.336 /
\plot  4.762 23.336  8.096 22.384 /
\setdots < 0.0952cm>
\plot  1.429 25.241  0.889 25.432 /
\plot  1.429 22.320  0.857 22.162 /
\plot  8.064 25.241  8.636 25.432 /
\plot  8.128 22.384  8.731 22.193 /
\setsolid
\circulararc 87.206 degrees from  8.096 21.431 center at  5.596 23.812
\plot  8.541 25.368  8.763 25.495 /
\put{$\tan\alpha=\frac{1}{m}$} [lB] at  4.953 21.431
\plot  8.604 22.257  8.858 22.193 /
\plot  0.984 25.368  0.730 25.464 /
\plot  1.016 22.225  0.730 22.098 /
\plot  4.826 24.035  4.762 24.289  4.699 24.035 /
\setdashes < 0.1270cm>
\plot  4.762 24.289  4.762 23.336 /
\setsolid
\plot  4.699 23.590  4.762 23.336  4.826 23.590 /
\setdashes < 0.1270cm>
\plot  4.762 24.289  7.430 24.289 /
\plot  8.096 23.812  9.049 23.812 /
\setsolid
\plot  8.795 23.749  9.049 23.812  8.795 23.876 /
\setdashes < 0.1270cm>
\plot  1.429 23.812  0.476 23.812 /
\setsolid
\plot  0.730 23.876  0.476 23.812  0.730 23.749 /
\setdashes < 0.1270cm>
\put{$\frac{1}{m}$} [lB] at  4.667 22.384
\put{$C_\infty$} [lB] at  9.334 24.574
\put{$\bo D^n$} [lB] at  3.334 21.241
\put{$\bo Q^n$} [lB] at  6.477 25.813
\put{$V_m$} [lB] at  1.429 24.479
\linethickness=0pt
\putrectangle corners at  0.451 26.211 and  9.334 21.184
\endpicture
$$
Consider the spaces of {\it exponentially decreasing holomorphic functions} on
these sets defined by
$$
\eqalign{\OM(V_m)&\equiv\{f\in\OO(U_m^n)|\,||f||_m<\infty\},\text{ where}\cr
||f||_m&\equiv\sup_{z\in\bo Q^n\cap V_m}|f(z)|e^{|z|/m}.}
\eqno(\OM)
$$
Taking the inductive limit of these spaces in the locally convex category
with respect to $m$, i.e., letting the neighbourhoods approach the real axis both
in distance and direction, we obtain the space of {\it rapidly decreasing
holomorphic functions} on $\bo D^n$:
$$
\PII\equiv\varinjlim_{m\rightarrow\infty}\OM(V_m),
\eqno{(\PII)}
$$
which is the space of test functions for the modified Fourier hyperfunctions, which
we can now define as their dual space $\eusm R(\bo D^n)\equiv\PIIp$.

{\bf Modified Fourier hyperfunctions as boundary values.}
Closer to the original ideas of Sato lies the definition of $\eusm R(\bo D^n)$
via {\it relative cohomology} --- which comprises the $n$-dimensional generalization
of the notion of boundary values of holomorphic functions. Consider
the set of {\it slowly increasing holomorphic functions} on an open subset
$U\subset\bo Q^n$ given by
$$
\OII(U)\equiv \left\{ f\in\OO(U)\,\Big|\,\sup_{z\in\bo Q^n\cap
      K}|f(z)|e^{-\varepsilon |z|}<\infty\ \forall\varepsilon>0,\
    K\Subset U\right\}.
\eqno{(\text{SI})}
$$
Then we can define a presheaf
$\{\eusm R(\Omega)\mid\Omega\subset\bo D^n\text{ open}\}$, by assigning to every
open subset $\Omega$ of $\bo D^n$ the $n$-th relative cohomology group with
support in $\Omega$ and coefficients in $\OII$:
$$
\eusm R(\Omega)\equiv H_\Omega^n(U,\OII),
$$
for any complex neighbourhood $U$ of $\Omega$ (see~\cite{MO} for a concise
introduction to relative cohomologies and their use in hyperfunction theory).
Let us just note aside, that in one dimension this definition reduces to simply
taking equivalence classes:
$\eusm R(\bo D^1)=\OII(\bo Q^1\setminus\bo D^1)/\OII(\bo Q^1)$. One finds with
these definitions:
\theo{}
The presheaf $\{\eusm R(\Omega)\}$ is a flabby sheaf.
Further, for any compact set $K$ in $\bo D^n$ we have
$\eusm R(K)=H_K^n(U,\OII)\simeq\PIIp|_K$. In particular
$\eusm R(\bo D^n)$ is isomorphic to the dual space of $\PII$.
\endtheo
See, e.g.,~\cite{15}, theorems~3.5 and~3.6. The second part is the fourier
hyperfunction equivalent of the famous Martineau-Harvey duality theorem, which
can be found in~\cite{MO}, theorem~6.5.1.

{\bf Euclidean Reconstruction using modified Fourier hyperfunctions.}
The general strategy of reconstruction used in~\cite{15} is the same as
the original one of Osterwalder-Schrader in the tempered case. Naturally, the
temperedness estimate (RE) is to be replaced by a weaker one in consistence
with the slowly increasing property (SI) of the boundary value representation
of $\PIIp$. Indeed after formulating axiomatic quantum field theory with
(modified) Fourier hyperfunctions\footnote{Actually, they use Fourier
hyperfunctions of a mixed type, which are ordinary Fourier hyperfunctions in
the spatial variables and modified only in the time variables, but this
shall not concern us here.}
 in~\cite{14} and~\cite{15}, Nagamachi-Mugibayashi
carry out the analytic continuation as outlined in section~2, and find for
the corresponding Schwinger functions an estimate of the form
$$
|\Sf_n(\xi)|\leq C^{(0)}_{n,\varepsilon}\cdot e^{\varepsilon|\xi|}\quad \forall
\varepsilon>0,
$$
on the non-coincident Euclidean region, see theorems~4.2 and~4.6 in~\cite{15},
which is there called {\it infra-exponential} estimate. Again, the real analytic
functions $\Sf_n$ admit an analytic continuation to cones $\Gamma^{(N)}$ of
increasing opening angle in the time variables and satisfiy a continued estimate
analoguously to (NE):
$$
|\Sf_n(\zeta)|\leq C^{(N)}_{n,\varepsilon}\cdot e^{\varepsilon|\zeta|},\
\forall\varepsilon>0,\ \zeta\in\Gamma^{(N)},
\eqno{(\text{NI})}
$$
see~\cite{15}, proposition~5.3. Now, for the continued $\Sf_n$ to define
a distribution on $\PII(\bo D^{4(n-1)})$ as wanted, one has to show that
the evaluation of $\Sf_n$ on any test function $f\in\PII(\bo D^{4(n-1)})$
makes sense. By definition ($\PII$) such a function will be in one of the spaces
building the inductive limit, i.e., we can take $f_m\in\OM(V_m)$, where
$V_m$ is a neighbourhood of $\bo D^{4(n-1)}$ of the same form as in ($V_m$), i.e.,
with finite opening angle above the real ``axis''. So, can we define the integral
$$
\Sf_n(f_m)\equiv
\int_{\gamma(N,m)} \Sf_n(\zeta)f_m(\zeta)d\zeta,
$$
for a certain integration contour $\gamma=\gamma(N,m)$ near $\bo D^{4(n-1)}$,
depending on the domains of holomorphy of both $f_m$ and $\Sf_n$ (namely
$\Gamma^{(N)}$)? Indeed such a contour exists as shown below:
$$
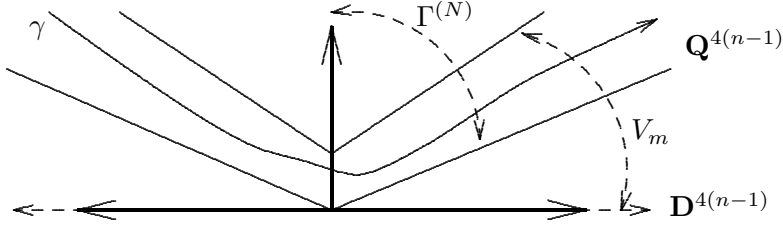
\beginpicture
\setcoordinatesystem units <.9cm,.9cm>
\linethickness=1pt
\setlinear
\linethickness= 0.500pt
\setdashes < 0.1270cm>
\setsolid
\plot  6.691 25.017  6.763 24.765  6.818 25.021 /
\setdashes < 0.1270cm>
\setsolid
\plot  5.007 26.574  4.762 26.480  5.022 26.448 /
\setdashes < 0.1270cm>
\circulararc 95.274 degrees from  6.763 24.765 center at  4.981 24.710
\linethickness= 0.500pt
\plot  8.096 23.812  9.049 23.812 /
\setsolid
\plot  8.795 23.749  9.049 23.812  8.795 23.876 /
\setdashes < 0.1270cm>
\linethickness=1pt
\setsolid
\plot  1.842 23.940  1.334 23.812  1.842 23.685 /
\putrule from  1.334 23.812 to  8.192 23.812
\plot  7.684 23.685  8.192 23.812  7.684 23.940 /
\linethickness= 0.500pt
\setdashes < 0.1270cm>
\plot  1.429 23.812  0.476 23.812 /
\setsolid
\plot  0.730 23.876  0.476 23.812  0.730 23.749 /
\setdashes < 0.1270cm>
\linethickness=1pt
\setsolid
\putrule from  4.762 23.812 to  4.762 26.289
\plot  4.890 25.781  4.762 26.289  4.635 25.781 /
\linethickness= 0.500pt
\plot  1.905 26.480  4.762 24.574 /
\plot  4.762 24.574  7.620 26.480 /
\linethickness= 0.500pt
\plot  0.381 25.718  4.762 23.812 /
\plot  4.762 23.812  9.334 25.718 /
\linethickness= 0.500pt
\setdashes < 0.1270cm>
\setsolid
\plot  8.679 24.074  8.668 23.812  8.800 24.038 /
\setdashes < 0.1270cm>
\setsolid
\plot  7.596 26.189  7.334 26.194  7.563 26.066 /
\setdashes < 0.1270cm>
\circulararc 91.356 degrees from  8.668 23.812 center at  6.838 24.352
\put{$\gamma$} [lB] at  0.681 26.194
\linethickness= 0.500pt
\setsolid
\plot  0.952 26.480      1.024 26.428
         1.095 26.377
         1.164 26.327
         1.232 26.278
         1.299 26.230
         1.365 26.183
         1.430 26.137
         1.493 26.091
         1.617 26.003
         1.736 25.919
         1.851 25.838
         1.962 25.760
         2.068 25.685
         2.171 25.614
         2.271 25.545
         2.366 25.480
         2.458 25.418
         2.547 25.358
         2.633 25.301
         2.715 25.247
         2.795 25.195
         2.871 25.146
         2.945 25.099
         3.016 25.054
         3.085 25.012
         3.152 24.972
         3.278 24.897
         3.396 24.831
         3.508 24.771
         3.614 24.718
         3.715 24.670
         3.780 24.644
         3.860 24.617
         3.950 24.591
         4.046 24.564
         4.142 24.540
         4.234 24.517
         4.315 24.496
         4.382 24.479
         4.456 24.457
         4.545 24.427
         4.645 24.394
         4.752 24.360
         4.859 24.329
         4.964 24.305
         5.060 24.290
         5.143 24.289
         5.242 24.307
         5.352 24.343
         5.471 24.392
         5.591 24.449
         5.710 24.510
         5.821 24.570
         5.919 24.625
         6.001 24.670
         6.065 24.706
         6.137 24.749
         6.215 24.798
         6.300 24.852
         6.388 24.910
         6.480 24.970
         6.574 25.033
         6.669 25.097
         6.763 25.161
         6.857 25.224
         6.949 25.285
         7.037 25.343
         7.121 25.397
         7.199 25.447
         7.270 25.490
         7.334 25.527
         7.455 25.592
         7.522 25.627
         7.595 25.664
         7.674 25.704
         7.760 25.746
         7.854 25.791
         7.956 25.840
         8.066 25.892
         8.186 25.948
         8.250 25.977
         8.317 26.008
         8.386 26.040
         8.458 26.073
         8.533 26.107
         8.610 26.143
         8.691 26.179
         8.775 26.218
         8.862 26.257
         8.953 26.298
         9.047 26.340
         9.144 26.384
        /
\plot  8.938 26.222  9.144 26.384  8.886 26.338 /
\put{$\bo Q^{4(n-1)}$} [lB] at  9.525 25.908
\put{$V_m$} [lB] at 8.8 24.765
\put{$\Gamma^{(N)}$} [lB] at  5.906 26.289
\put{$\bo D^{4(n-1)}$} at 10 23.9
\linethickness=0pt
\putrectangle corners at  0.356 26.594 and  9.525 23.766
\endpicture
$$
It exists if the opening angle of $\Gamma^{(N)}$ around $\imath\bo D^{4(n-1)}$
is larger than that of the given $V_m$, that is, after {\it finitely many steps}
$N(m)$ of analytic continuation. The integral $\Sf_n(f_m)$ is then clearly finite
by the characterization of the test function ($\OM$) and the estimate (NI), see
theorem~3.15 in~\cite{15}.

The point is, that this result can be reached after finitely many steps for
every given test function. this makes it simply {\it unnecessary} to control
the growth of the constant $C^{(N)}_{n,\varepsilon}$ in $n$ and thereby in $N$.
So what we have found is that Euclidean reconstruction of modified Fourier
hyperfunctions is possible {\it without growth conditions}.

It may seem as if
the {\it modification} of the Fourier hyperfunctions, namely the introduction
of test functions of exponential decrease in domains of finite opening angle
over $\bo D^n$ is just as {\it ad hoc} as the linear growth condition. But let
me point out that in the sense of boundary values such a condition is dictated
by the choice of compactification of $\bo C^n$, and the radial compactification,
which gives equal importance to real and imaginary coordinates naturally leads
to ($V_m$) and ($\OM$). We will compare this with ordinary Fourier hyperfunctions
below.
%
% -------------------------------------------
\head 4.~Outlook\endhead
The two sorts of Euclidean reconstruction discussed above represent two
extremes: The strong linear growth condition leading to the reconstruction
of very ``regular'' --- namely tempered --- distributions, and the absence of
growth conditions resulting in a very ``singular'' class of reconstructed
functions --- namely hyperfunctions --- which can for example contain
singularities like $exp(1/x)$. This phenomenon seems to be more of a
fundamental than of a technical type. As is well known from the variants of
the Payley-Wiener-Schwartz theorem one has as a rule of thumb that weaker growth
conditions correspond to more singular distributions as boundary
values\footnote{Remember that in the Osterwalder-Schrader procedure the bound
in $n$ was used to get a bound in $N$, i.e., essentially in the distance from
the real axis.}, see, e.g., the discussion at he beginning of chapter~8
of~\cite{KA}.

This naturally raises the question if one could find intermediate cases between
these extremes. Florin Constantinescu has since long proposed to use
{\it ultradistributions} in that connection,
see~\cite{21},~\cite{22},~\cite{23}, and the introductory text~\cite{24},
but I want to point out another direction which could be taken.

{\bf Reconstruction of ordinary Fourier hyperfunctions.}
If one compactifies only $\bo R^n$ by adding the points at infinity to
get $\bo D^n$ and uses $\bo D^n+i\bo R^n$ as the base space from which
boundary values on the real axis are taken as hyperfunctions, one is led
to the definition of ordinary Fourier hyperfunctions, also termed
{\it Fourier hyperfunctions
of type I} in~\cite{14}, where the quantum field theory is formulated
with the latter. On the side of the test function spaces the difference
to the modified ones lies in the form of the complex neighbourhoods of
the real axis in the equivalents of definitions ($V_m$) and ($\OM$). They
have to be chosen as parallel stripes along the real axis
with certain width in the imaginary direction:
$$
U_{1;m}\equiv\{z\in\bo C\mid|\Im z|<1/m\}.
$$
One can the proceed to build the space $\PI$ of test functions for
Fourier hyperfunctions as an inductive limit as in ($\PII$). One immediately
sees $\PII\subset\PI$ since the elements of $\PII$ have to satisfy growth
conditions,and in fact to be analytic on larger domains as that of $\PI$.
By that, $\PIIp$ is a larger, i.e. ``more singular,'' class of distributions.

Trying to prove an Euclidean reconstruction theorem yielding
boundary values in $\PIp$ leads to the same difficulties as in the tempered
case. Let us consider a fixed test function $f\in\OM(V_{1;m})\subset\PI$.
For this $f$, the integral $\Sf(f)$ with the analytic continuation of
$\Sf_n$ can only be carried out partially on the finite integration
path $\gamma$ shown in the sketch below.
$$
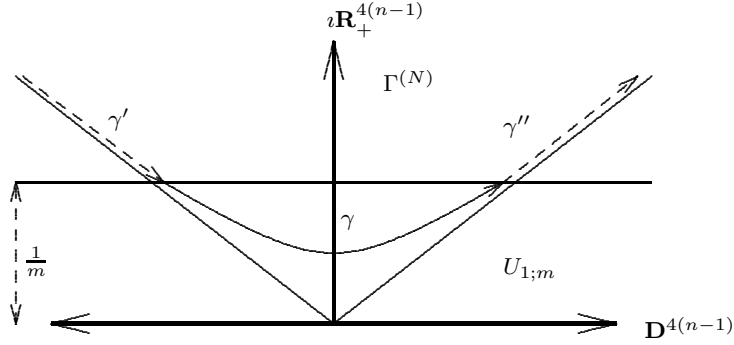
\beginpicture
\eightpoint
\setcoordinatesystem units <.9cm,.9cm>
\linethickness=1pt
\putrule from  4.762 22.384 to  4.762 26.194
\plot  4.890 25.686  4.762 26.194  4.635 25.686 /
\linethickness= 0.500pt
\plot  0.476 25.718  4.762 22.384 /
\plot  4.762 22.384  9.049 25.718 /
\putrule from  0.476 24.289 to  9.049 24.289
\plot  0.413 22.638  0.476 22.384  0.540 22.638 /
\setdashes < 0.1270cm>
\plot  0.476 22.384  0.476 24.289 /
\setsolid
\plot  0.540 24.035  0.476 24.289  0.413 24.035 /
\setdashes < 0.1270cm>
\plot  0.572 25.718  2.476 24.289 /
\setsolid
\plot  2.235 24.390  2.476 24.289  2.311 24.492 /
\setdashes < 0.1270cm>
\plot  7.048 24.289  8.858 25.718 /
\setsolid
\plot  8.698 25.510  8.858 25.718  8.620 25.610 /
\plot  2.476 24.289      2.588 24.223
         2.697 24.160
         2.801 24.100
         2.902 24.043
         3.000 23.988
         3.095 23.936
         3.186 23.886
         3.275 23.839
         3.360 23.794
         3.443 23.751
         3.523 23.711
         3.600 23.673
         3.675 23.638
         3.748 23.605
         3.818 23.574
         3.886 23.545
         3.952 23.518
         4.016 23.493
         4.138 23.449
         4.254 23.413
         4.364 23.385
         4.469 23.363
         4.570 23.348
         4.667 23.339
         4.762 23.336
         4.858 23.339
         4.955 23.348
         5.056 23.363
         5.161 23.385
         5.271 23.413
         5.387 23.449
         5.509 23.493
         5.573 23.518
         5.639 23.545
         5.707 23.574
         5.777 23.605
         5.850 23.638
         5.925 23.673
         6.002 23.711
         6.082 23.751
         6.165 23.794
         6.250 23.839
         6.339 23.886
         6.430 23.936
         6.525 23.988
         6.623 24.043
         6.724 24.100
         6.828 24.160
         6.937 24.223
         7.048 24.289
        /
\plot  6.862 24.105  7.048 24.289  6.797 24.215 /
\linethickness=1pt
\plot  1.460 22.511  0.952 22.384  1.460 22.257 /
\putrule from  0.952 22.384 to  8.572 22.384
\plot  8.064 22.257  8.572 22.384  8.064 22.511 /
\put{$\imath\bo R^{4(n-1)}_+$} [lB] at  4.667 26.384
\put{$\gamma^\prime$} [lB] at  1.715 25.051
\put{$\Gamma^{(N)}$} [lB] at  5.429 25.527
\put{$\gamma^{\prime\prime}$} [lB] at  7.048 24.955
\put{$\gamma$} [lB] at  4.858 23.717
\put{$U_{1;m}$} [lB] at  7.048 23.050
\put{$\bo D^{4(n-1)}$} [lB] at  8.954 22.193
\put{$\frac{1}{m}$} [lB] at  0.572 23.146
\linethickness=0pt
\putrectangle corners at  0.451 26.714 and  9.074 22.111
\endpicture
$$
This shows that, as in the tempered case, $\Sf_n$ defines a functional
on the test function space only after infinitely many steps of analytic
continuation, calling again for growth conditions in $N$.

Maybe the situation is not so hopeless. One could try to approximate the test
function $f$ by certain functions $f_k$ such that the integrals $\Sf_n(f_k)$
exist on the whole path $\gamma'\circ\gamma\circ\gamma''$. Such an
approximation could possibly consist of quasi-analytic
extensions
of $f$ out of the stripe $U_{1;m}$ as frequently used by H\"ormander,
see~\cite{H\"O}.
For such extensions one has $L^2$-estimates, and one could hope to calculate
what I would call the {\it functional error}
$$
 R(N,n;m,k)\equiv\left|\int_{\gamma^\prime}+
    \int_{\gamma^{\prime\prime}}\Sf_n(\zeta)f_m(\zeta)d\zeta\right|.
$$
which loosely speaking measures the distance from $\Sf_n$ to a functional
on the whole space $\PI$. Trying then to control $R(N,n;m,k)$ in $m$ and $k$
should make it possible by the same procedures as used by Osterwalder-Schrader
to pose a condition on the original Schwinger functions
like the linear growth condition that make $R(N,n;m,k)$ vanish in the limit
$N\rightarrow\infty$ and thus is a sufficient condition for Euclidean
reconstruction of Fourier hyperfunctions.

{\bf Intermediate spaces of hyperfunctions.}
One could think of constructing a sequence of spaces continuously mediating
between $\PI$ and $\PII$ by simply modifying the geometry.
Set for $0\leq\rho\leq1$
$$
\eqalign{%
    U_{\rho;m}&\equiv\{z\in\bo C|\,|\Im z|<(1+|\Re z|)^\rho/m\},\cr
    \PR&\equiv\varinjlim_{m\rightarrow\infty}\OM(V_{\rho;m}),}\eqno{(\PR)}
$$
where $V_{\rho;m}$ is the closure of $U^n_{\rho;m}$ in $\bo Q^n$ as in
($V_m$). The definition of the inductive limit can be easily shown to be
independent of the exact form of $U_{\rho;m}$, i.e., wether one takes
$(1+|\Re z|)^\rho/m$, $(1+|\Re z|^\rho)/m$ or $\{(1+|\Re z|)/m\}^\rho$
as the upper bound for the imaginary part of $z$ in the definition.
Then one finds the spaces of (modified) Fourier hyperfunctions
as limit cases $\PI=\eusm P_0$ and $\PII=\eusm P_1$ in the sequence
$\PR\subset\eusm P_{\rho'}$, $\rho<\rho'$, of {\it intermediate spaces}.
The $\PR$ can also be used as test function spaces for Fourier hyperfunctions:
\lem{}
The Fourier transformation is a topological isomorphism on $\PR$.
\endlem
This can be easily shown following the proof of proposition~3.2 in~\cite{13}
with minor modifications.

As the intersection of $U_{\rho;m}$ with $\Gamma^{(N)}$ becomes larger as
$\rho$ increases the error term $R(N,n;m,k)$ should decrease, and
it would be very tempting to get by the proceeding outlined above a continuous
sequence of growth conditions which pose no restriction in the limit case
$\eusm P_1$ and tend to the --- yet unknown --- growth condition needed
for reconstruction of distributions in $\PIp$ in the limit
$\rho\rightarrow0$.

{\bf The scheme of Gelfand-Shilov spaces.}
In~\cite{GS}, which is volume II of the famous series of books on functional
analysis and distribution theory, Gelfand and Shilov introduced a general
classification of test function spaces for distributions. They classified
test functions by growth order in coordinates and derivatives by defining
$$
\eqalign{%
\SA &\equiv
\{f\in\eusm C^\infty(\bo R)\mid |x^k f^{(q)}(x)|\leq C_q A^k k^{k\alpha}\},\cr
\SB &\equiv
\{f\in\eusm C^\infty(\bo R)\mid |x^k f^{(q)}(x)|\leq C_k B^q q^{q\beta}\},\cr
\SAB &\equiv
\{f\in\eusm C^\infty(\bo R)\mid |x^k f^{(q)}(x)|\leq
  C  A^k B^q k^{k\alpha}  q^{q\beta} \}.}
\eqno{(\SAB)}
$$
Within this scheme are contained many known examples of test function spaces
such as $\eusm D$ and $\eusm S$, and they are set into relation by the well
examined properties of the $\SAB$ and their behaviour with
respect to varying indices. The sketch below, whose details we will now briefly
discuss, presents the $\SAB$-scheme.

First of all, the spaces with only one index are limit cases of the general ones,
i.e., we can extend the definition of $\SAB$ to infinite values
of $\alpha$ and $\beta$ and find
$\SB=\eusm S_\infty^\beta= \lim_{\alpha\rightarrow\infty}\SAB$,
$\SA=\eusm S^\infty_\alpha= \lim_{\beta\rightarrow\infty}\SAB$, and finally
one recognizes $\eusm S_\infty^\infty$ to be identical with the space of
rapidly decreasing infinitely differentiable functions $\eusm S$. One also finds
the test function space $\eusm D$ of $\eusm C^\infty$-functions with compact
support in the upper left corner, $\eusm D=\eusm S^\infty_0$, and its Fourier
transform in the lower right: $\eusm{FD}=\eusm S_\infty^0$.

This is, as one could already have guessed from $(\SAB)$, a general phenomenon,
showing the duality of the indices $\alpha$ and $\beta$ with respect to Fourier
transformation: $\eusm F\SAB=\eusm S_\beta^\alpha$. As a special case we find that
the spaces $\eusm S_\alpha^\alpha$ on the diagonal are closed under Fourier
transformation.
$$
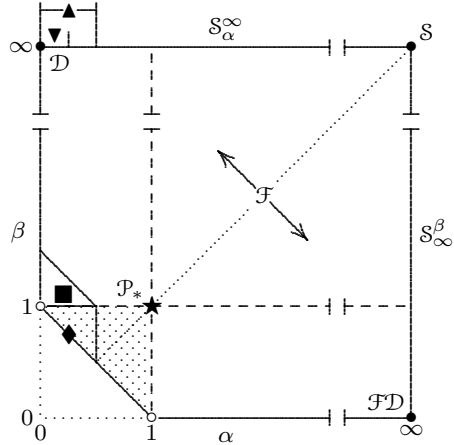
\beginpicture
\eightpoint
\setcoordinatesystem units <0.9mm,0.9mm>
\setplotarea x from -10 to 60, y from -10 to 60
\plot 16 0 39 0 /
\plot 41 0 50 0 50 39 /
\plot 50 41 50 50 41 50 /
\plot 39 50 0 50 0 41 /
\plot 49 39 51 39 /
\plot 49 41 51 41 /
\plot 39 49 39 51 /
\plot 41 49 41 51 /
\plot 0 39 0 16 /
\plot 39 1 39 -1 /
\plot 41 1 41 -1 /
\plot -1 39 1 39 /
\plot -1 41 1 41 /
\plot 14.5 0.5 0.5 14.5 /
\setdots <3pt>
\plot 0 14.5 0 0 14.5 0 /
\setdots <2pt>
\plot 7.5 7.5 28 28 /
\plot 32 32 50 50 /
\setdashes <3pt>
\plot 1 15 39 15 /
\plot 41 15 50 15 /
\plot 15 1 15 39 /
\plot 15 41 15 50 /
\setsolid
\plot 14 39 16 39 /
\plot 14 41 16 41 /
\plot 39 14 39 16 /
\plot 41 14 41 16 /
\put {$\circ$} [cc] at 15 0
\put {$\circ$} [cc] at 0 15
\put {$\bullet$} [cc] at 50 0
\put {$\bullet$} [cc] at 50 50
\put {$\bullet$} [cc] at 0 50
%\put {$\bullet$} [cc] at 15 15
\put {$0$} [ct] at 0 -1
\put {$0$} [rc] at -1 0
\put {$1$} [ct] at 15 -1
\put {$1$} [rc] at -1 15
\put {$\infty$} [ct] at 50 -1
\put {$\infty$} [rc] at -1 50
\put {$\eusm S$} [lb] at 51 51
\put {$\eusm F$} [cc] at 30 30
\put {$\eusm D$} [lt] at 1 49
\put {$\eusm{FD}$} [rb] at 49 1
\put {$\PI$} [rb] at 14 16
\put {$\alpha$} [ct] at 25 -2
\put {$\beta$} [rc] at -2 25
\put {$\eusm S_\alpha^\infty$} [cb] at 25 51
\put {$\eusm S^\beta_\infty$} [lc] at 51 25
\arrow <2mm> [0.25,0.75] from 31 29 to 36 24
\arrow <2mm> [0.25,0.75] from 29 31 to 24 36
\put {$\bigstar$} [cc] at 15 15
\put {$\blacktriangle$} [cc] at 3.8 55
\put {$\blacklozenge$} [cc] at 3.75 11.25
\put {$\blacksquare$} [cb] at 3 15.5
\plot 7.5 15 0 22.5 /
\plot 1 15 7.5 15 7.5 7.5 /
\put {$\blacktriangledown$} [cc] at 1.875 51.5
\plot 0 50 0 56 /
\plot 3.75 50 3.75 52 /
\plot 7.5 50 7.5 56 /
\plot 0 55 3 55 /
\plot 5 55 7.5 55 /
\setlinear
\setshadegrid span <2pt>
\vshade 0 15 15 15 0 15 /
\endpicture
$$

As the index $\beta$ controls the derivatives of a function
$f\in\SAB$, it can also control the convergence of its Taylor series
and by that the analytic continuation of $f$, if it exists. In fact,
for $\beta\leq1$, every such $f$ can be analytically continued to some
stripe around the real axis depending on $B$ and characteristics of
$f$. The stripes grow in width like $1/(eB)$ as $B$ decreases and
especially for $\beta<1$, i.e., below the dashed horizontal line,
$\SAB$ consists of entire functions. This also means that the
functions in $\SAB$ with $\alpha<1$ have Fourier transforms which are
entire functions. See also~\cite{26} for a characterization of $\SAB$
in terms of Fourier transforms.

It is apparent form the definitions that $\SAB$ is contained in the
intersection of $\SA$ and $\SB$, but the other inclusion is also true as
was shown in~\cite{27}, so that we have $\SAB=\SA\cap\SB$.
We also see that the Gelfand-Shilov spaces become smaller with decreasing
indices, as sharper conditions are imposed on the functions. Indeed, $\SAB$
can be shown to be trivial, i.e., contain only constants, for $\alpha+\beta<1$,
as well as are $\eusm S_0^1$ and $\eusm S^0_1$.

As already mentioned, models of interacting quantum fields can only be
formulated in more singular classes of distributions than the tempered one,
due to the singularities of the correlation functions which can be deduced
from the constructive ingredients of the model, the Lagrangean, commutation
relations, covariance, and so on. The suitable test function spaces for certain
models were classified in the Gelfand-Shilov scheme in~\cite{25}:

{\parindent=1.5cm
 \item{$\bullet$} The {\it two-dimensional dipole field}, which is a solution of
$\square^2\varphi=0$, has a two-point function which is a distribution on
$\SA$, for $\alpha<1/2$, marked by $\blacktriangle$.
 \item{$\bullet$} The {\it vertex operator} $\mathord{:}\exp(ig\varphi)\mathord{:}$
of this field
is even more singular, as it is an operator valued distribution over
$\SA$, $\alpha<1/4$, which is marked by $\blacktriangledown$.
 \item{$\bullet$} $\mathord{:}\exp(ig\varphi)\mathord{:}$
for the dipole field in 4 spacetime
dimensions is --- in the time variables --- a distribution
over $\SAB$, $\alpha+\beta<3/2$, $\alpha<1/2$, the area marked by $\blacksquare$.
\par}

Most important for our discussion, also (modified) Fourier hyperfunctions can be
arranged into the scheme. The test function space $\PI$ of Fourier hyperfunctions
was shown in~\cite{14} to be isomorphic to $S_1^1$, the point which is marked by
$\bigstar$. For $\PII$, only a dense subspace could be found, see~\cite{15},
lemma~2.1 and proposition~2.2., with $\eusm S^{3/4}_{1/4}$ (or $\eusm S^{1/4}_{3/4}$
since $\PII$ is closed under Fourier transformation), which is marked by
$\blacklozenge$.

It should be possible to arrange the intermediate spaces $\PR$ into the
$\SAB$-frame in a similar manner. This can give a first hint if growth
conditions are needed for an Euclidean reconstruction on $\PRp$, since the
fact that $\PII$ contains a dense subspace of {\it entire} functions seems
to play a r\^{o}le in this respect. So far we have found:
\lem{}
For all $\alpha$, $\beta$ $\in (0,1)$ with $\alpha+\beta\geq 1$
and $0\leq\rho\leq1$ we have
$\SAB\subset \PR$ for $\alpha\leq  (1-\beta)/\rho$,
and the topology of $\SAB$ is stronger than that induced by $\PR$ in this case.
In particular $\eusm S_\alpha^\alpha\subset\PR$ for $\alpha\leq (1+\rho)^{-1}$.
\endlem
\proof{}
We use an alternative definition of $\SAB$ as an inductive limit of locally
convex spaces, which may be found in~\cite{15}, section~2:
$$
\eqalign{
\SAB&=\varinjlim_{n\rightarrow\infty} \eusm T_\alpha^{\beta,n},\cr
T_\alpha^{\beta,n}&\equiv \{ f \text{ entire}\mid ||f||_{(\alpha,\beta);n}<\infty\},
\quad\text{where}\cr
||f||_{(\alpha,\beta);n}&\equiv
\sup_{z\in\bo C} |f(z)|e^{n^{-1}|\Re z|^{1/\alpha}+ n|\Im z|^{1/(1-\beta)}}.
}
$$
Now it suffices to show that for every $f\in\eusm T_\alpha^{\beta,n}$ we find
some $m$ such that $||f||_{\rho;m}\leq C ||f||_{(\alpha,\beta);n}$, where
$||.||_{\rho;m}$ is the norm introduced in $\OM(V_{\rho;m})$ by formula ($\OM$).
Since then we have continuous embeddings
$T_\alpha^{\beta,n}\hookrightarrow\OM(V_{\rho;m})$ $\hookrightarrow\PR$ and by that
of $\SAB$ in $\PR$.
To estimate the norms it is enough to estimate the exponents, i.e., to show
$$
n^{-1}|\Re z|^{1/\alpha}+ n|\Im z|^{1/(1-\beta)}\geq
m^{-1}|z|, \quad\forall z\in U_{\rho;m},
$$
for some $m$. It is enough to verify this condition on the boundary
of the domain $U_{\rho;m}$, where the imaginary part becomes maximal,
i.e., $|\Im z|=m^{-1}(1+\Re z)^\rho$. With that and using
$|z|\leq|\Re z|+|\Im z|$, $a=\alpha^{-1}$, $b=(1-\beta)^{-1}$,
$x\equiv|\Re z|$, we have
only to show
$$
n^{-1}x^a\geq nm^{-b}(1+x)^{\rho b} + m^{-1}x+m^{-1}(1+x)^\rho,
$$
for $x>0$ and a suitable choice of $m$. Since $a$, $b>1$, $0\leq\rho\leq1$,
this is always achieveable
for sufficiently large $m$ if $a\geq\rho b$, which is equivalent to
$\alpha\leq (1-\beta)/\rho$, which completes the proof.
\endproof

One would like to know if the largest of the subspaces of $\PR$ lying on
the diagonal, i.e., $\eusm S_\alpha^\alpha$, $\alpha=(1+\rho)^{-1}$ is dense
in $\PR$, but this has yet to be examined.

{\bf Related Work.}
In~\cite{28} Yury M. Zinoviev established a full equivalence between a
certain formulation of Euclidean and the Wightman field theory by inventing
a new inversion formula for the Fourier-Laplace transformation for tempered
distributions. But he had to impose an additional condition on the Euclidean
side which he calls {\it weak spectral condition}, and whose physical meaning
is somewhat unclear to me.

A further extension of distribution theory building on the
Glefand-Shilov classification but exceeding it, was proposed as an
extension of ultradistribution and hyperfunction theory suitable for
the formulation of gauge quantum field theories in~\cite{29}. This
extension corresponds to spaces $\SAB$ with $\beta<1$ in momentum
space in accordance with the classification of some models, which was
discussed above, in configuration space. He proves a sort for
localizability of distributions on these spaces and other technical
tools needed for quantum field theory. This extension is applied to
the rigorous definitionof Wick-Ordered entire functions of free fields
in~\cite{30}.

To conclude let us say that the question of the `right' test function
spaces for interacting quantum fields is still unresolved, but at
least {\it modified} hyperfunctions for which, as we have shown above
$\eusm S^{3/4}_{1/4}$ is a suitable test function space, are a
reasonable one candidate. This is because these hyperfunctions allow
to include all physical models discussed above, and equally important
the beautiful theoretical machinery of hyperfunctions is at hand for
them. On the other hand one should not be surprised if it needs
further generalizations as in~\cite{29,30} if one wants to go over
from `toy--models' to `real--world problems'.
%
% -------------------------------------------
\head Acknowledgements \endhead
I would like to thank Florin Constantinescu for his continuing support.
%
%%%%%%%%%%%%%%%%%%%%%%%%%%%%%%%%%%%%%%%%%%%%%%%%%%%%%%%%%%%%%%%%%%%%%%%%%
%%                               REFERENCES                            %%
%%%%%%%%%%%%%%%%%%%%%%%%%%%%%%%%%%%%%%%%%%%%%%%%%%%%%%%%%%%%%%%%%%%%%%%%%
%

%
%%%%%%%%%%%%%%%%%%%%%%%%%%%%%%%%%%%%%%%%%%%%%%%%%%%%%%%%%%%%%%%%%%%%%%%%%
%%                               AUTHOR'S ADDRESS                      %%
%%%%%%%%%%%%%%%%%%%%%%%%%%%%%%%%%%%%%%%%%%%%%%%%%%%%%%%%%%%%%%%%%%%%%%%%%
%
\bigskip
{\flushpar
\eightpoint
%\rec{{November 11, 1997}}
\addr{Andreas U. Schmidt}
\addr{Fachbereich Mathematik}
\addr{Johann Wolfgang Goethe--Universit{\"a}t}
\addr{D--60054 Frankfurt am Main}
\addr{{\it{e--mail:}}}                      % --- my e-mail does not fit in
\addr{\it aschmidt\@math.uni--frankfurt.de}}% --- a single line !!
\enddocument